\documentclass[prl,aps,twocolumn,amsmath,amssymb,showpacs]{revtex4}
\usepackage{graphicx}% Include figure files
\usepackage{dcolumn}% Align table columns on decimal point
\usepackage{bm}% bold math

\begin{document}
\title{Hysteresis and Anisotropic Magnetoresistance in Antiferromagnetic $Nd_{2-x}Ce_xCuO_{4}$}
\author{X. H. Chen$^1$}
\altaffiliation{Corresponding author} \email{chenxh@ustc.edu.cn}
\author{C. H. Wang$^1$, G. Y. Wang$^1$, X. G. Luo$^1$, J. L. Luo$^2$, G. T. Liu$^2$ and N. L.
Wang$^2$} \affiliation{1. Hefei National Laboratory for Physical
Science at Microscale and Department of Physics, University of
Science and Technology of China, Hefei, Anhui 230026, People's
Republic of China\\ 2. Beijing National Laboratory for Condensed
Matter Physics, Institute of Physics, Chinese Academy of Science,
Beijing 100080, People's Republic of China}

\begin{abstract}
The out-of-plane  resistivity ($\rho_c$) and  magnetoresistivity
(MR) are studied in antiferromangetic (AF) $Nd_{2-x}Ce_xCuO_{4}$
single crystals, which have three types of noncollinear
antiferromangetic spin structures. The apparent signatures are
observed in $\rho_c(T)$  measured at the zero-field and 14 T at
the spin structure transitions, giving a definite evidence for the
itinerant electrons directly coupled to the localized spins. One
of striking feature is an anisotropy of the MR with a fourfold
symmetry upon rotating the external field (B) within ab plane in
the different phases, while twofold symmetry at spin reorientation
transition temperatures. The intriguing thermal hysteresis in
$\rho_c(T,B)$ and magnetic hysteresis in MR are observed at spin
reorientation transition temperatures.
\end{abstract}
\vskip 15 pt


 \narrowtext

\maketitle

High-$T_c$ superconductivity occurs in cuprates when doping
introduces sufficient holes or electrons into the $CuO_2$ planes.
It is generally believed that the pairing necessary for
supercoductivity involves the interplay between the doped charges
and the AF spin correlation. In this sense, the study of lightly
doped, insulating AF state is important because the density of the
carriers can be sufficiently low that the interaction between them
is small relative to their interaction with the $Cu^{+2}$ spins.
Many intriguing and anomalous phenomena show up in lighly doped AF
cuprates due to the strong coupling between charges and magnetic
order of the $Cu^{2+}$ spins\cite{thio,ando1,ando2,ando3}.

 In the hole doped cuprates, the Neel order is rapidly suppressed by doped
hole, resulting in a "spin-glass" state \cite{kastner} and a
strong tendency to form spin-charge textures or "stripes"
\cite{tranquada}. However, the long-rang AF order in
electron-doped $Nd_{2-x}Ce_xCuO_4$ persists to much larger x
($\geq$0.12) \cite{tokura}, and coexists with superconductivity
for even the optimal doping material (x=0.15) with Tc=25 K
\cite{yamada}. In addition, the $Cu^{+2}$ spins order in an AF
collinear structure for the parent compounds (such as: $La_2CuO_4$
and $YBa_2Cu_3O_6$) of hole-doped cuprates
\cite{vaknin,tranquada1}, while in AF noncollinear structure for
that of electron-doped cuprates: $Pr_2CuO_4$ and $Nd_2CuO_4$
\cite{skanthakumar,sumarlin}. All spins point either parallel or
antiparallel to a single direction in AF collinear structure,
while the spins in adjacent layers are orthogonal in AF
noncollinear structure. Magnetic-field induced a transition from
noncollinear to collinear spin arrangement in adjacent $CuO_2$
planes for lightly electron-doped $Pr_{1.3-x}La_{0.7}Ce_xCuO_4$
with x=0.01 crystals affects significantly both the in-plane and
out-of-plane resistivity \cite{ando3}. In $Nd_2CuO_4$, the
$Cu^{2+}$ spins order in three phases with two different AF
noncollinear spin structures and experience two reorientation
phase transitions
\cite{skanthakumar1,matsuda,skanthakumar,skanthakumar2} as shown
in Fig.1. Such reorientation phase transition is absent in
$Pr_2CuO_4$\cite{sachidanandam}.

Magnetoresistance (MR) provides new insight into the coupling
between the charges and the background magnetism. Previous
experiments \cite{thio,ando2,ando3} have demonstrated that
\begin{figure}[b]
\includegraphics[width=8cm]{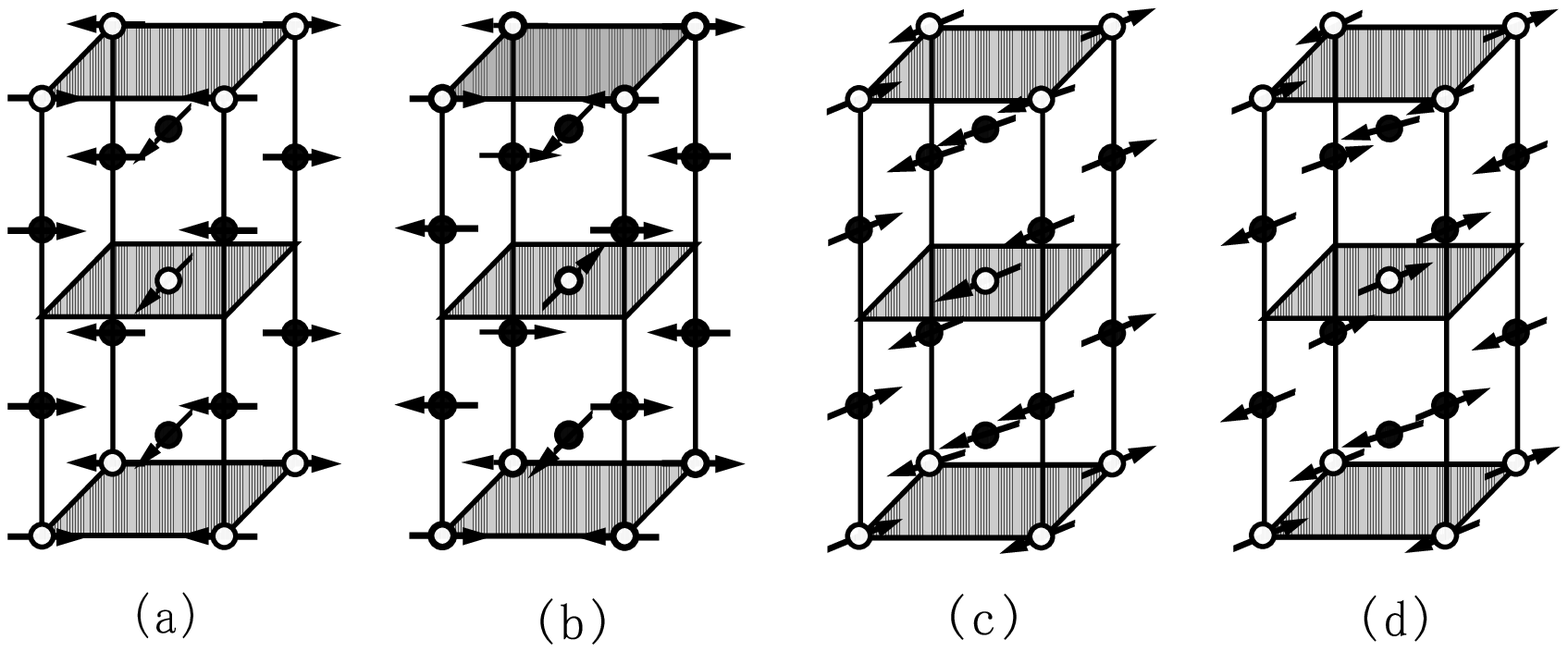}\vspace{-2mm}
\caption{\label{fig:epsart} Spin structure models for the AF
noncollinear structure at zaro field and the relative orientation
of spins for the AF collinear structure at high field. (a)
Noncollinear Phase I (75 $<$ T $<$ 275 K) and Phase III (T$<$30
K); (b) Phase II (30$<$ T$<$75 K); (c) Collinear Phase (type-I and
III) induced by the field along the [110] from Phase I and III;
(d) Collinear Phase (type-II) from Phase II. Here the open circles
are Cu ions and the solid ones Nd ions.}\vspace{-2mm}
\end{figure}
out-of-plane resistivity is sensitive to the interlayer magnetic
order of the $Cu^{2+}$ spins. This is particularly valuable
because, as shown in this work, the spin-flop or flip transition
occurs at fields in which magnetization measurements are
difficult. In this letter, we systematically studied out-of-plane
MR and angular dependence of out-of-plane MR for lightly doped
$Nd_{2-x}Ce_xCuO_4$. It is found that $\rho_c(T)$ is surprisingly
sensitive to the spin reorientation, giving a definite evidence
for the itinerant electrons directly coupled to the localized
spins. A thermal hysteresis in $\rho_c(T)$ at field and magnetic
hysteresis in MR are observed. Another striking feature is the MR
anisotropy with a fourfold symmetry in different AF spin
structures, while with a twofold symmetry at the spin
reorientation temperatures.

The NCCO single crystals were grown by flux method over a wide
range of Ce concentration $0\leq x \leq 0.13$. The actual Ce
concentration was determined by inductively coupled plasma
spectrometry (ICP) analysis experiements, and by the
energy-dispersive x-ray analysis (EDX) using a scanning electron
microscopy, respectively. All samples were annealed in flowing
helium with purity of 99.999\% for 10 hours at 900 $^oC$ to remove
the interstitial oxygen. The resistivity and magnetoresistance
were performed in Quantum Design PPMS systems.

\begin{figure}[t]
\includegraphics[width=8.5cm]{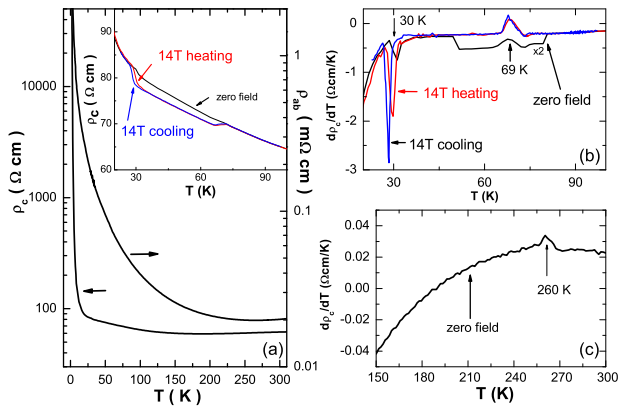}\vspace{-6mm}
\caption{\label{fig:epsart} (a) In-plane and out-of-plane
resistivity as a function of temperature for
$Nd_{1.975}Ce_{0.025}CuO_4$ crystal. (b)-(c) Temperature
derivative of out-of-plane resistivity. In (b), the zero field
data are multiplied by 2 between 50 to 80 K. }\vspace{-5mm}
\end{figure}

The magnetic properties in $R_2CuO_4$ (R=Nd, Pr) are mainly
dependent on the coupling between Cu and R magnetic subsystem
which exhibits a large single-ion anisotropy \cite{sachidanandam}.
Unlike $Pr_2CuO_4$, in $Nd_2CuO_4$ the spin reorientation
transition takes place due to a competition between various
interplanar interactions which arises because of the rapid
temperature dependence of the Nd moment below about 100 K
\cite{sachidanandam}. The Cu spins first order in the noncolinear
AF structure [phase I: Fig.1a] below $T_{N1}$=275 K. On further
cooling, the Cu spins reorder into the noncollinear structure at
$T_{N2}$=75 K [phase II: Fig.1b], and at $T_{N3}$=30 K the Cu
spins experience another reorientation into phase III which has
the same noncollinear order as phase I \cite{endoh,skanthakumar}.
As shown in Fig.1, the Cu and Nd moments along c-axis are parallel
in phase I and III, while antiparallel in phase II
\cite{petitgrand}. A magnetic field applied within ab planes
induces a transition from the noncollinear to collinear AF
structure with a spin flop. Fig.1(c) and (d) show the collinear AF
structures transformed from the noncollinear structures shown in
Fig.1(a) and (b) at the B applied along [110] direction.

Figure 2a shows temperature dependence of in-plane ($\rho_{ab}$)
and out-of-plane ($\rho_c$) resistivity for the crystal
$Nd_{1.975}Ce_{0.025}CuO_4$. $\rho_{ab}(T)$ shows the insulating
behavior in the whole temperature range, while $\rho_{c}(T)$ shows
a weak metallic behavior above 130 K, and a weak insulating
behavior with decreasing temperature, and a diverging below 30 K.
At a first glance, no anomaly is detected at $T_{N1}$, $T_{N2}$
and $T_{N3}$ in the zero field $\rho_{ab}(T)$ and $\rho_c(T)$.
However, a derivative plot for out-of-plane resistivity shown in
Fig.1(b) and Fig(c) helps to observe the anomalies at the
different $T_N$. As shown in Fig.1(b) and 1(c), the clear peaks
are observed at $T_{N2}$$\sim$69 K and $T_{N3}$$\sim$30 K
 for the spin reorientation transition, while a weak peak shows up at
 about 260 K for the AF order. Compared to $Nd_2CuO_4$, the $T_{N1}$ and $T_{N2}$
 slightly decreases. It suggests that doping of Ce suppresses the AF order and spin reorientation at
 $T_{N2}$, while does not affect the $T_{N3}$ remarkably.
  At $T_{N2}$ and $T_{N3}$, the width of the spin
 reorientation transition is very narrow (less than 10 K), indicating
 the high quality of our crystals. It should be pointed out that
 no anomaly is observed in ab-plane resistivity, even in its derivative.
 It suggests a strong coupling  between the charge and spin degree of
 freedom, and that out-of-plane resistivity is more sensitive to the spin
 structure than in-plane resistivity. It should be pointed out that the anomalies
 shown in $d\rho_{c}/dT$ can be observed only for the $Nd_{2-x}Ce_xCuO_4$
 crystals with x$<$0.06. In order to investigate the effect of
 collinear AF structure transition on $\rho_{c}(T)$, the
 $\rho_{c}(T)$ is measured at the B of 14 T along [110] direction in the
 heating and cooling process. As shown in the inset of Fig.2a, a remarkable feature
 is observed in $\rho_{c}(T)$ at $T_{N2}$ and $T_{N3}$, and the transition from the type-I collinear
 structure to the type-II leads to a decrease in $\rho_{c}$.
  Very sharp peaks show up in a derivative plot ($\rho_c(T)/dT$) at  $T_{N2}$ and
 $T_{N3}$, respectively. The peak position at $T_{N2}$ remains
 unchanged, while at $T_{N3}$ obviously shifts to low
 temperature. A intriguing feature is that a hysteresis behavior
 at 14 T is clearly observed at $T_{N3}$ and the peak temperature difference between heating and cooling process
 is about 1.5 K,
 while a hysteresis behavior can be ignored at $T_{N2}$.
These results give the definite evidence for the itinerant
electrons directly coupled to the localized spins. The similar
hysteresis behavior
 cannot be observed at zero field. Therefore, the hysteresis is
 induced by the external field.

 \begin{figure*}[t]
\includegraphics[width=\textwidth]{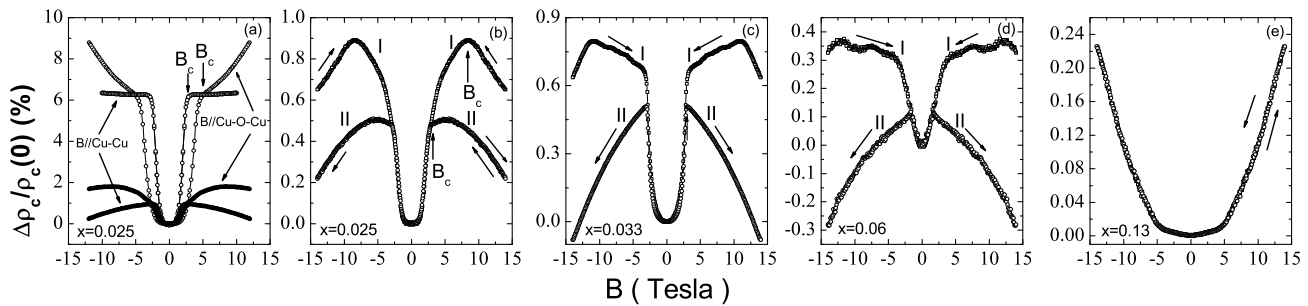}\vspace{-14mm}
\caption{\label{fig:wide} (a) The isothermal MR  for x=0.025 with
the B along Cu-O-Cu and Cu-Cu direction at 20 K (open) and 40 K
(solid), respectively; (b)-(e) The isothermal MR as a function of
the B at 50 K in FC and ZFC process with B$\|$Cu-Cu direction for
the $Nd_{2-x}Ce_xCuO_4$ crystals with x=0.025, 0.033, 0.06 and
0.13, respectively.}\vspace{-5mm}
\end{figure*}

Figure 3(a) shows the isothermal MR  for x=0.025 with the B along
Cu-O-Cu and Cu-Cu direction at 20 K and 40 K, respectively. The MR
behavior shown in Fig.3(a) is similar to that observed in
$Pr_{1.3-x}La_{0.7}Ce_xCuO_4$ with x=0.01 crystals by Lavrov et
al\cite{ando3}. As explained by Lavrov et al., the MR behavior
arises from the spin origin and is closely related to the
noncollinear-collinear transition induced by B. The steplike
increase of resistivity corresponds to the noncollinear-collinear
transition with increasing the field up to critical field $B_c$,
above the $B_c$ (in collinear structure) the MR shows different
behavior. As shown in Fig.3(a), the Cu-Cu direction is easy axis
in the collinear spin structure with relatively small $B_c$.
 Figure 3(b)-(e) shows the isothermal MR
 at 50 K with B applied within ab-plane Cu-Cu direction
  for the $Nd_{2-x}Ce_xCuO_4$ crystals with x=0.025, 0.033, 0.06 and 0.13.
   An intriguing result is that magnetic hysteresis of MR is observed in the
 for x=0.025, 0.033 and 0.06 crystals. The magnetic field dependence of isothermal MR
 shows two branches. The branch of larger MR is obtained with field-cooled (FC) process, that is, the B of 14 T or -14 T
 is applied within ab-plane at 290 K, then the sample is cooled to 50 K with B,
 and the isothermal MR is measured with decreasing B.  The branch of the smaller MR is got
 with increasing B from zero to 14 T, then with decreasing B the MR shows the same behavior as shown in
 Fig.3(b). The isothermal MR shows the same behavior as the smaller MR branch
 in zero-field cooled (ZFC) process. It suggests that the isothermal MR is
 strongly dependent on the B applied history. In $Pr_2CuO_4$, no similar magnetic hysteresis of the MR is observed.
While the difference of their magnetic structure is the absence of
the spin reorientation in $Pr_2CuO_4$. Therefore, the unique
feature of the magnetic hysteresis in MR is closely related to the
spin reorientation. As pointed out by Sachidanandam et
al.\cite{sachidanandam}  the spin reorientation transition
originates from the competition between various interplanar
interactions because of the rapid temperature dependence of the Nd
moment below 100 K. It is possible that the magnetic field has an
effect on various interplanar interactions.  So the different
effect of the field on the interplanar interactions  exists in the
different collinear spin structure,  and leads to the different MR
behavior between the ZFC and FC process. No magnetic hysteresis
observed in ab-plane MR supports this explanation. No magnetic
hysteresis is observed in Fig.3(e) for the x=0.13 crystal. This
may be due to two possibilities: (1) antiferromagnetic order is
completely suppressed by doping, or the $T_{N1}$ is below 50 K;
(2) the spin reorientation temperature $T_{N2}$ is suppressed to
be less than 50 K with doping, so that no spin orientation occurs
above 50 K as the case of $Pr_2CuO_4$. There exist two important
differences between the MR branch I and II. First is the MR
behavior above $B_c$ and the sign of the anomalous MR, which is
always positive for branch I, while negative at high fields for
branch II. Second, the critical field $B_c$ for the noncollinear
to collinear spin structure transition is larger in branch I than
that in branch II. Which could originate from the effect of the B
on the interplanar interactions in the FC process enhances the
critical field for the noncollinear-collinear spin transition. It
should be pointed out that the hysteresis in MR is not observed
below $T_{N3}$, and only can be observed at temperature between
$T_{N2}$ and $T_{N3}$.

 In order to make out effect of spin reorientation
transition on the MR, the MR as a function of angle at different
temperatures is studied. The angular dependence of the MR was
determined by rotating the sample under a fixed field of 12 T
within ab-plane. Figure 4 shows the evolution of the MR with angle
between B and [100] (Cu-O-Cu) direction for the 0.025 crystal. The
MR is always positive for all temperatures. A striking feature is
that the MR shows a strong anisotropy with fourfold symmetry in
different AF phases, while twofold symmetry around $T_{N2}$ and
$T_{N3}$. The similar anisotropy with \textbf{\emph{d-wave-like}}
symmetry has been observed in $Pr_{1.3-x}La_{0.7}Ce_xCuO_4$ with
x=0.01 crystal \cite{ando3}. The fascinating MR oscillations shown
in Fig.4 has been explained to arise from the relative orientation
of spins with respect to the crystal axes because the spin
structure always stays collinear at high fields. The total energy
does not change due to the interplane pseudo-dipolar interactions
when the spin sublattices of the adjacent $CuO_2$ planes rotate in
opposite directions \cite{petitgrand1,plakhty,sachidanandam}. Such
continuous spin rotation can be induced by B because the spins
gradually rotate toward a configuration perpendicular to the B at
high fields. Therefore, the hard and easy spin axes are tuned by
the field. A intriguing feature is the evolution of MR diagram
with temperature. With decreasing temperature, the fourfold
oscillations in type-I collinear phase are replaced by a twofold
sine wave like feature at $T_{N2}$; consequently a new fourfold
oscillations show up in type-II phase, and are replaced by another
twofold symmetric wave like feature across $T_{N3}$; with further
cooling fourfold symmetric oscillations develop. The MR diagram is
fairly
\begin{figure}[t]
\includegraphics[width=8cm]{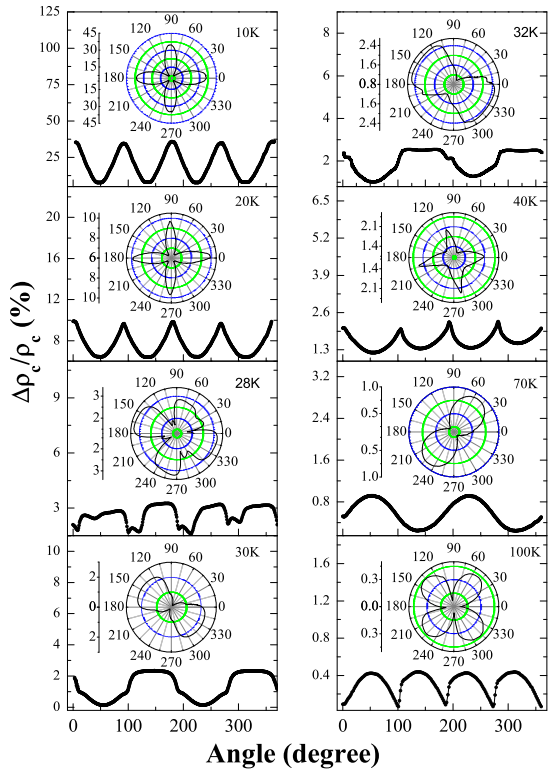}\vspace{-14mm}
\caption{\label{fig:epsart} The out-of-plane MR as a function of
angle for $Nd_{1.975}Ce_{0.025}CuO_4$ crystal at different
temperatures and polar plot of the same data.}\vspace{-5mm}
\end{figure}
symmetric in type-I and type-III collinear phase, while is
asymmetric in type-II phase. The MR diagram rotates by 45 $^oC$ in
the type-II collinear phase relative to the type-I collinear
phase, and the MR diagram with twofold symmetry at $T_{N2}$ and
$T_{N3}$ rotates by 90 $^oC$ each other. Surprisingly, this is
quite consistent with the spin reorientation from Phase I to II
(all the Cu spins rotated by $90^o$ about the c-axis). However, it
is strange that the MR diagram does not change from type-II to
type-III collinear structure. So the maximum MR is observed with B
along Cu-O-Cu direction in type-II and type-III collinear
structure, while with B along Cu-Cu direction in type-I phase.
This is consistent with the magnetic hysteresis in MR observed in
Phase II and no magnetic hysteresis in MR observed in Phase III.
These results should be closely related to the Nd and Cu ion
interaction, so that the MR shows different behavior for the same
Cu spin order in phase I and III. It suggests that the MR effects
are quite sensitive to the differences in the different collinear
spin structures.

It should be pointed out that no MR anomaly and no hysteresis
behavior in MR and in $\rho_c(T)$ are observed when a c-axis
aligned field is applied, consistent with no transformation from
noncollinear to collinear spin structure for such field
orientation\cite{matsuura}. Compared to the lightly doped
Pr-Ce-Cu-O material, the MR behavior is similar to each other due
to the transition from noncollinear to collinear spin structure.
However, the unique feature is the thermal hysteresis in
$\rho_c(T)$ and the magnetic hysteresis in MR. The thermal
hysteresis at spin reorientation transition cannot be observed at
zero field $\rho_c(T)$, suggesting that the field has effect on
the interplanar interactions since the spin reorientation
transition arises from the competition of the various interplanar
interactions. It is the effect of B on various interplanar
interactions to lead to the FC MR behavior different from the ZFC
MR case as shown in Fig.3. In addition, the evolution of the MR
diagram with the temperature shown in Fig.4 is consistent with the
spin structure transition at $T_{N2}$. The maximum MR appears with
B along Cu-O-Cu direction in type-II collinear phase, while with B
along Cu-Cu direction in type-I collinear phase. It suggests that
the hard and easy spin axes are different in type-I and -II
collinear spin structures. Which could be the origin for the
different MR behavior between FC and ZFC processes shown in Fig.3.
The MR diagram does not change across $T_{N3}$, so that no
magnetic hysteresis in MR is observed. But a remarkable thermal
hysteresis is observed at $T_{N3}$. It suggests that the thermal
hysteresis observed in Fig. 2 has a different origin from the
magnetic hysteresis in MR shown in Fig.3.

In summary, the transport properties and the MR behavior  are
systematically studied in antiferromagnetic $Nd_{2-x}Ce_xCuO_{4}$.
The transport properties is very sensitive to the subtle changes
of the spin structure. We give a direct evidence for the itinerant
electrons directly coupled to the localized spins. The thermal
hysteresis in $\rho_c(T,B)$ and the magnetic hysteresis in MR are
found. The hysteresis arises from the effect of the field on the
interplanar interactions, such as: coupling between Cu and Nd
ions, and the different hard and easy spin axes in the collinear
spin structures.

 Upon preparing this Letter, we became aware of a
similar hysteresis observed in neutron scattering and MR
experiments for $Nd_{2-x}Ce_xCuO_{4}$ \cite{li}.

 We would like thank Pengcheng Dai, X. G. Wen, Tao Xiang and Qianghui Wang for helpful discussions.
  This work is supported by the grant from the National Natural Science Foundation of China
 and the Knowledge Innovation Project of Chinese Academy of Sciences.\\

\end{document}